\documentclass[aps,prb,twocolumn,shortbibliography,superscriptaddress,article]{revtex4-1}
\usepackage{epsfig}
\usepackage{epstopdf}
\usepackage{amsmath}
\usepackage{amsfonts}
\usepackage{amssymb}
\usepackage{hyperref}
\usepackage{bm}
\usepackage{makecell}
\usepackage{rotating}
\usepackage{hyperref}
\usepackage{multirow}
\usepackage{graphicx}
\usepackage{booktabs}
\usepackage{siunitx}
\usepackage{makecell}
\usepackage{color,soul}
\usepackage{ulem}
\usepackage{tabularx}
\usepackage{booktabs}
\usepackage{array}

\usepackage{graphicx}% Include figure files
\usepackage{dcolumn}% Align table columns on the decimal point
\usepackage{bm}% bold math
\usepackage{color}

\usepackage{tikz,xcolor,hyperref}

\definecolor{lime}{HTML}{A6CE39}
\DeclareRobustCommand{\orcidicon}{%
	\begin{tikzpicture}
	\draw[lime, fill=lime] (0,0)
	circle [radius=0.16]
	node[white] {{\fontfamily{qag}\selectfont \tiny ID}};
	\draw[white, fill=white] (-0.0625,0.095)
	circle [radius=0.007];
	\end{tikzpicture}
	\hspace{-2mm}
}

\foreach \x in {A, ..., Z}{%
	\expandafter\xdef\csname orcid\x\endcsname{\noexpand\href{https://orcid.org/\csname orcidauthor\x\endcsname}{\noexpand\orcidicon}}
}

\begin{document}

\title{Defect Geometry Selects Polar \\ and Anomalous Hall Phases in Two-Dimensional Altermagnets}

\author{Xujia Gong\orcidX}
\email{xgong@magtop.ifpan.edu.pl}
\affiliation{International Research Centre Magtop, Institute of Physics, Polish Academy of Sciences,
Aleja Lotnik\'ow 32/46, PL-02668 Warsaw, Poland}

\author{Amar Fakhredine\orcidF}
\affiliation{Institute of Physics, Polish Academy of Sciences, Aleja Lotnik\'ow 32/46, 02668 Warsaw, Poland}

\author{Hosein Alavi-Rad\orcidY}
\affiliation{Department of Electrical Engineering, Lan.C., Islamic Azad University, Langarud, Iran}

\author{Mahyar Hassani-Vasmejani\orcidZ}
\affiliation{Department of Computer Engineering, Lan.C., Islamic Azad University, Langarud, Iran}

\author{Xing Ming}
\email{mingxing@glut.edu.cn}
\affiliation{College of Physics and Electronic Information Engineering, Key Laboratory of Low-dimensional Structural Physics and Application, Education Department of Guangxi Zhuang Autonomous  Region, Guilin University of Technology, Guilin 541004, People's Republic of China}

\author{Xiangang Wan}
\affiliation{National Laboratory of Solid State Microstructures and Department of Physics, Nanjing University, Nanjing 210093, China}

\author{Carmine Autieri\orcidB}
\email{autieri@magtop.ifpan.edu.pl}
\affiliation{International Research Centre Magtop, Institute of Physics, Polish Academy of Sciences,
Aleja Lotnik\'ow 32/46, PL-02668 Warsaw, Poland}

\author{Meysam Bagheri Tagani\orcidA}
\email{mtagani@magtop.ifpan.edu.pl}
\affiliation{Department of Physics, University of Guilan, P. O. Box 41335-1914, Rasht, Iran}
\affiliation{International Research Centre Magtop, Institute of Physics, Polish Academy of Sciences, Aleja Lotnik\'ow 32/46, PL-02668 Warsaw, Poland}

\author{Sahar Izadi Vishkayi\orcidS}
\email{izadi.123@gmail.com}
\affiliation{International Research Centre Magtop, Institute of Physics, Polish Academy of Sciences, Aleja Lotnik\'ow 32/46, PL-02668 Warsaw, Poland}

\date{\today}
\begin{abstract}
Point defects in altermagnets can create phases absent in the pristine host by selectively breaking crystal symmetries. Combining symmetry analysis, first-principles calculations, and Hamiltonian modeling, we identify how point impurities modify the altermagnetic phase. Using the pristine $d$-wave altermagnetic monolayer V$_2$Se$_2$O as a testbed, we identify three distinct classes of impurities: those that preserve spin-momentum locking, those that induce a hybrid-parity state associated with Edelstein spin conversion, and those that produce a metallic ferrimagnetic state with an anomalous Hall effect. We further discuss the robustness of two-dimensional altermagnets against point impurities. 
Results for other two-dimensional systems, such as Mn$_4$N$_2$ and 2H-FeBr$_3$, reveal the same symmetry-based control across distinct lattices and parent spin harmonics, establishing defect geometry as a general route for engineering spin textures and transport properties. 
\end{abstract}

\pacs{}

\maketitle

\textit{Introduction.—}
Altermagnets exhibit non-relativistic spin-momentum locking (SML) in momentum space with even parity, reconciling the negligible stray fields and ultrafast dynamics of antiferromagnets with the potential for spintronic devices
\cite{vsmejkal2022emerging,vsmejkal2022beyond,Hayami19,bai2024altermagnetism,
song2025altermagnets,bose2026symmetry,shao2021spin}. Spin-resolved photoemission has resolved the
even-parity SML, while transport and dichroic imaging have experimentally established altermagnetic properties
\cite{krempasky2024altermagnetic,reimers2024direct,ding2024large,
amin2024nanoscale,badura2025observation,badura2025observation}. Its spin-group symmetry\cite{PhysRevX.12.021016} further permits
unconventional spin, optical, and anomalous Hall responses 
\cite{jungwirth2026symmetry,zhao2026emergent,Benny26,PhysRevB.111.054442,Fakhredine26,
hirakida2025multipoleanalysisspincurrents,reichlova2024observation,mcclarty2024landau,zhou2025manipulation,liu2025anomalous,liu2025d}. In the extension to the relativistic case, the even parity persists in the presence of inversion symmetry and extends to all spin components\cite{Fakhredine26}. When the inversion symmetry is broken, the Rashba terms introduce p-wave contributions in the Hamiltonian\cite{10.1039/d6mh00357e}, making a system with a hybrid odd-even parity\cite{leon2025strainenhancedaltermagnetismca3ru2o7,luo2026unconventionalmagnetismsymmetryclassification}. The interplay with the even-wave SML determines if the system has defined parity with a hybrid or ill-defined parity\cite{gong2026symmetryprotectednodalplanesaccidental,luo2026unconventionalmagnetismsymmetryclassification}. 

Most symmetry classifications, however, assume an ideal periodic crystal\cite{tian2026symmetry,cheong2025altermagnetism,bai2025anomalous,wang2025electric}.
Real materials contain vacancies, antisites, and defect complexes that may remove precisely the operations responsible for magnetic compensation and spin splitting\cite{hu2024antisite,lnn6-hg4z}. Recent extensions to magnetic supercells\cite{JaeschkeUbiergo2024Supercell}, quasicrystals\cite{
Chen2025Quasicrystalline,Li2025Hyperspatial,Shao2025Classification}, synthetic antiferromagnets\cite{gallardo2026syntheticaltermagnetismcrystallimit} and impurity models\cite{gondolf2025local,k36v-91br,Das_2026,vina2025building,mavani2026competing} show that altermagnetic signatures do not require a primitive translational cell.
Whether realistic defects merely broaden an existing altermagnetic state or can systematically select new magnetic symmetries remains unresolved, especially in two-dimensional materials.
In this Letter, we address this question in two-dimensional systems by combining magnetic-symmetry analysis, density-functional theory (DFT), and Hamiltonian modeling. All vacancies, divacancies, and antisites considered here fall into three classes determined by the global magnetic symmetry that survives structural relaxation. Defect chemistry controls formation energy, charge redistribution, and response magnitude; symmetry determines whether the system remains a centrosymmetric altermagnet with preserved SML, becomes a polar spin--charge-active altermagnet with hybrid odd and even wave SML, or loses the symmetry protection of its SML, becoming a metallic ferrimagnet.

\begin{figure*}
    \centering
    \includegraphics[width=1\linewidth]{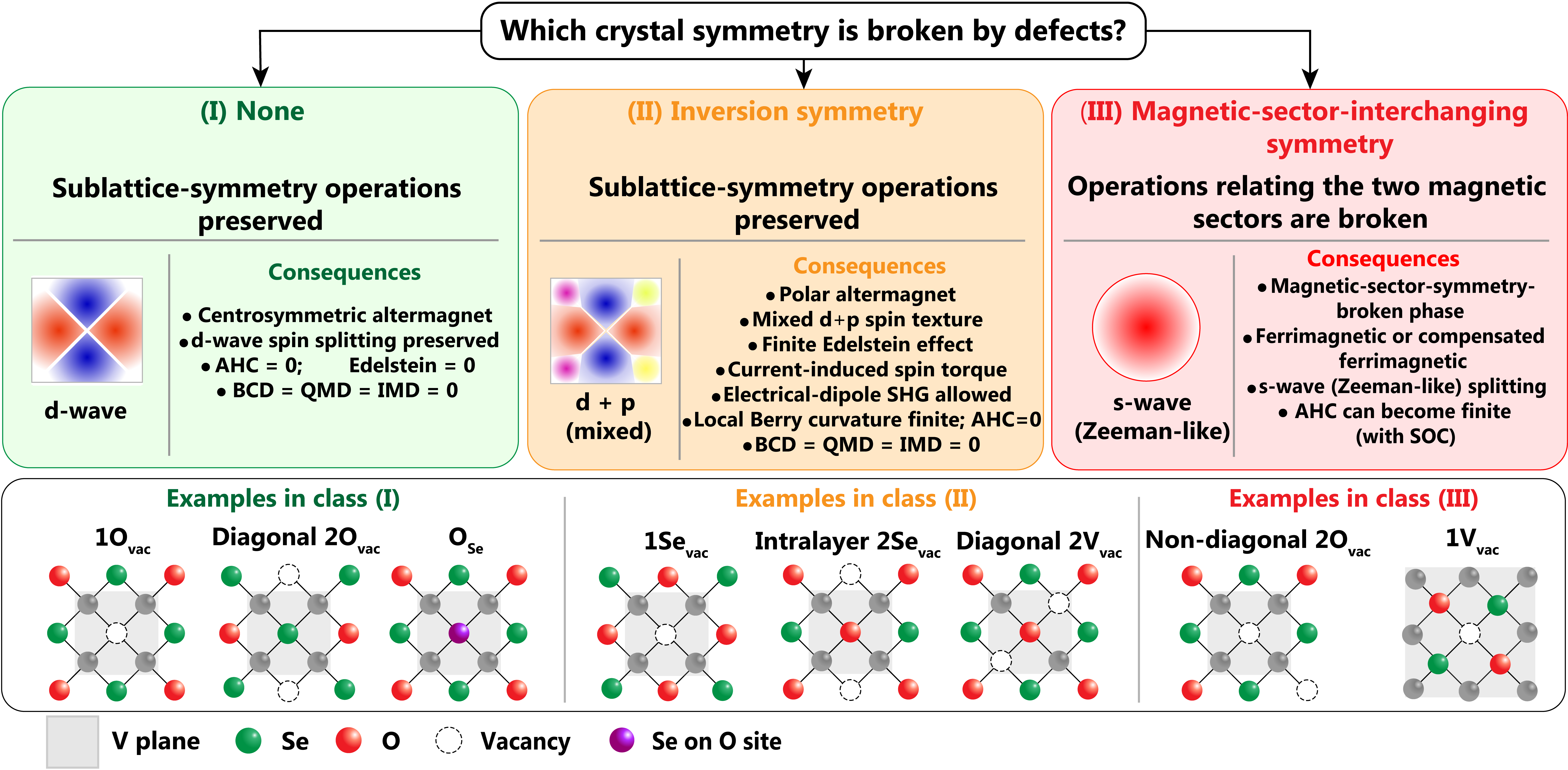}
    \caption{
\textbf{Symmetry-selective defect engineering in monolayer
V$_2$Se$_2$O.}
Defects follow three symmetry classes. 
Class I preserves inversion and a sector-relating symmetry operation and
retains the $d$-wave altermagnet, where the uniform Edelstein
response and intrinsic AHC vanish. Class II removes inversion
but preserves magnetic-sector equivalence, producing a polar altermagnet
with a mixed $d+p$ SML, finite Edelstein response, finite local Berry
curvature but zero intrinsic AHC; indeed, the surviving $C_{2z}$ symmetry forbids
purely in-plane BCD, QMD, and IMD responses. Class III removes the operation relating the magnetic sectors, strongly deforms the protected $d$-wave SML, and can permit intrinsic Hall transport in the presence of SOC.
Representative defects are shown for each class.
}
    \label{Scheme} % Figure 1
\end{figure*}

\textit{Symmetry-selected defect phases.—}
We consider the altermagnetic V$_2$Se$_2$O as a testbed material.
Pristine V$_2$Se$_2$O belongs to space group $P4/mmm$ and, for the
pure altermagnetic state considered here, to the magnetic space group
$P4^{\prime}/mm^{\prime}m$ (BNS No.~123.342)~\cite{sun2025symmetry,singh2025v2se2o,ma2021multifunctional,ahmad2026large,xu2026chemical,zhu2026altermagnetic}, with magnetic
point group $4^{\prime}/mm^{\prime}m$. 
% 4-5 citations more on monolayer of V$_2$Se$_2$O. 
The oppositely polarized V atoms are related by a rotational symmetry rather than by a primitive translation or inversion, producing the characteristic $B_{1g}$, or $d_{x^2-y^2}$-wave SML\cite{hayami2020bottom,sun2025optically}. 
% Preservation of the complete tetragonal parent group is not required: the compensated altermagnetic state survives whenever the relaxed supercell retains an operation that continues to relate the two magnetic sectors. The residual group then determines which momentum harmonics remain allowed.

Once impurities are introduced into the system, we identify three distinct classes, organized according to their residual magnetic symmetries, whose properties are summarized in Fig.~\ref{Scheme}. Chemically distinct defects belong to the same class when they preserve the same symmetry operations, whereas different arrangements of the same defect species can give rise to qualitatively different states.
Class I defects lower the symmetry of the system, but they preserve inversion and a sector-relating symmetry, retaining the pristine $d$-wave altermagnetic state. Class II defects remove inversion while preserving magnetic-sector equivalence, allowing spin--orbit coupling (SOC) to add an odd-parity SML together with the $d$-wave SML. Class III defects remove the symmetry operation relating the magnetic sectors, rendering them inequivalent. Consequently, the system becomes a metallic ferrimagnet, with $s$-wave symmetry as the lowest-order wave symmetry in $k$-space.
The corresponding response hierarchy follows from the magnetic point groups of the relaxed structures, rather than from the minimal model\cite{liu2026symmetry}. Class I forbids both a uniform Edelstein response and intrinsic anomalous Hall conductivity (AHC). The polar Class II groups permit a transverse Edelstein response and an associated current-induced exchange torque.
Their surviving $C_{2z}$ rotation, however, forbids conventional purely in-plane second-order charge transport, including Berry-curvature-dipole (BCD), quantum-metric-dipole (QMD), and inverse-mass-dipole (IMD) contributions
\cite{sodemann2015quantum,Zhang2023Symmetry}. Local Berry curvature is
symmetry-allowed, but its Brillouin-zone integral vanishes. Class II is
therefore spin--charge active yet intrinsically Hall silent. Class III can support AHC when the residual magnetic point group permits the Hall
pseudovector and SOC supplies the required interband mixing\cite{vsmejkal2020crystal,takahashi2025elasto}. Complete
space-group assignments and tensor constraints are given in the
Supplemental Material.

\begin{figure*}
    \centering
    \includegraphics[width=0.99\linewidth]{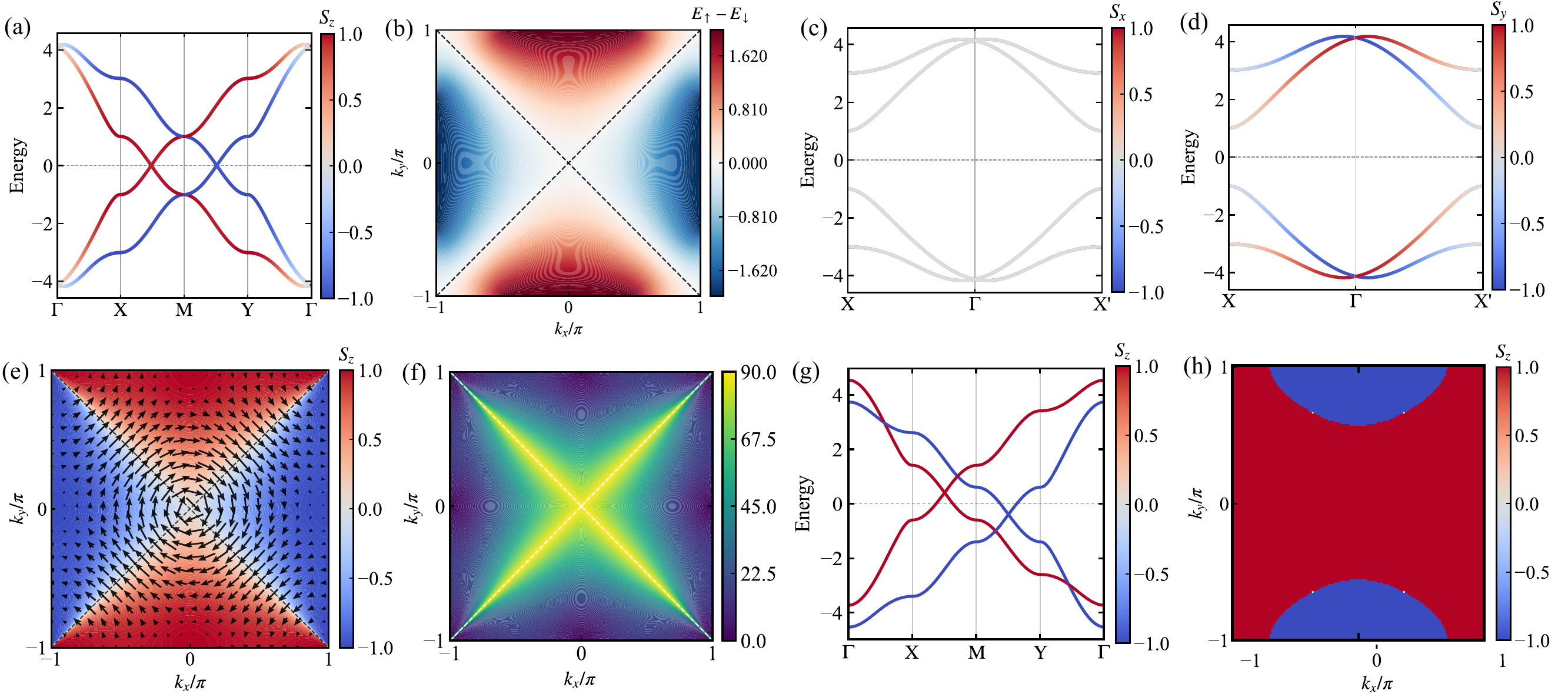}
    \caption{
\textbf{Minimal-model fingerprints of the three defect-selected phases.}
(a,b) Class I centrosymmetric altermagnet: spin-resolved band structure
and four-lobed $B_{1g}$ splitting. (c,d) Class II polar altermagnet:
bands projected onto $S_x$ and $S_y$,
respectively. Among the in-plane components, $S_x$ is symmetry suppressed
whereas $S_y$ is finite. (e,f) Mixed $d+p$ SML and corresponding
canting angle. (g,h) Class III band structure and momentum-space spin
texture, showing the emergence of ferrimagnetic phase after
magnetic-sector equivalence is removed. Numerical
details are given in the Supplemental Material.
}
\label{Fig2}
\end{figure*}

\textit{Model Hamiltonian.—}
We represent the two magnetic sectors by Pauli matrices
$\tau_i$ and spin by $\sigma_i$. The common centrosymmetric altermagnetic Hamiltonian is
\cite{bagheri2026quantum,roig2024minimal}:
\begin{equation}
H_{\rm AM}(\mathbf{k})
=
\xi(\mathbf{k})\tau_0\sigma_0
+
d_1(\mathbf{k})\tau_x\sigma_0
+
d_3(\mathbf{k})\tau_z\sigma_0
+
\Delta\tau_z\sigma_z ,
\label{eq:main_H_AM}
\end{equation}
where $\xi(\mathbf{k})$ is the single-band energy dispersion,
$d_1=4t\cos(k_x/2)\cos(k_y/2)$ hybridizes the two bands,
$d_3=2t_d(\cos k_x-\cos k_y)$ is the $B_{1g}$ SML, and
$\Delta$ is the staggered exchange field. To leading order in $\Delta$,
the non-relativistic spin-splitting is:
\begin{equation}
\delta E_{\eta}^{\rm AM}(\mathbf{k})
\simeq
2\Delta\eta
\frac{d_3(\mathbf{k})}
{\sqrt{d_1^2(\mathbf{k})+d_3^2(\mathbf{k})}} \quad \eta=\pm1
\label{eq:main_dwave_splitting}
\end{equation}
which is a $k_y^2-k_x^2$ SML near $\Gamma$; the exact spectrum and nodal planes are derived in the Supplemental Material.

Broken inversion permits the Rashba SOC in the form of: 
\begin{equation}
H_{\rm P}(\mathbf{k})
=
\left[
\alpha_x\sin k_y\,\sigma_x
-
\alpha_y\sin k_x\,\sigma_y
\right]\tau_0 .
\label{eq:main_H_P}
\end{equation}
The tetragonal polar phases impose
$\alpha_x=\alpha_y\equiv\alpha_{\rm R}$, whereas the orthorhombic phase
allows independent coefficients. Projection onto a hybridized band gives
the effective spin field
\begin{equation}
\mathbf{h}_{\eta}(\mathbf{k})
=
\left(
\alpha_x\sin k_y,\,
-\alpha_y\sin k_x,\,
\Delta\mathcal{G}_{\eta}(\mathbf{k})
\right),
~
\mathcal{G}_{\eta}
=
\eta\frac{d_3}{\sqrt{d_1^2+d_3^2}} .
\label{eq:main_effective_field}
\end{equation}
The polar phase therefore combines an even-in-momentum, sign-changing
out-of-plane exchange field with an odd-in-momentum in-plane winding. Its
bands remain reciprocal in energy, while their in-plane spin polarization
reverses between opposite momenta. Along $\Gamma$--X, symmetry imposes
$S_x=0$ while $S_y$ and $S_z$ may remain finite; along $\Gamma$--Y, the
complementary constraint applies.

For the representative Class III model, we use:
\begin{equation}
H_{\rm III}(\mathbf{k})
=
H_{\rm AM}(\mathbf{k})
+
M\tau_0\sigma_z
+
\lambda_{\rm I}\sin k_x\sin k_y\,\tau_y\sigma_z .
\label{eq:main_H_ClassIII}
\end{equation}
The uniform exchange component shifts the spin-splitting according to
$\delta E_{\eta}^{\rm III}=2M+\delta E_{\eta}^{\rm AM}$ and deforms the
four-lobed texture toward a predominantly ferrimagnetic polarization.
The parameter $M$ characterizes the uniform component of the effective
exchange potential.
The final term is a representative complex intersector SOC used to expose
the anomalous Hall mechanism: it produces local Berry curvature, whereas $M$ shifts the symmetry-related sectors and can prevent cancellation of their occupied contributions. The exact Class III
spectrum and the analytical cancellation relation at $M=0$ are given in
the Supplemental Material.

\begin{figure*}
    \centering
    \includegraphics[width=0.99\linewidth]{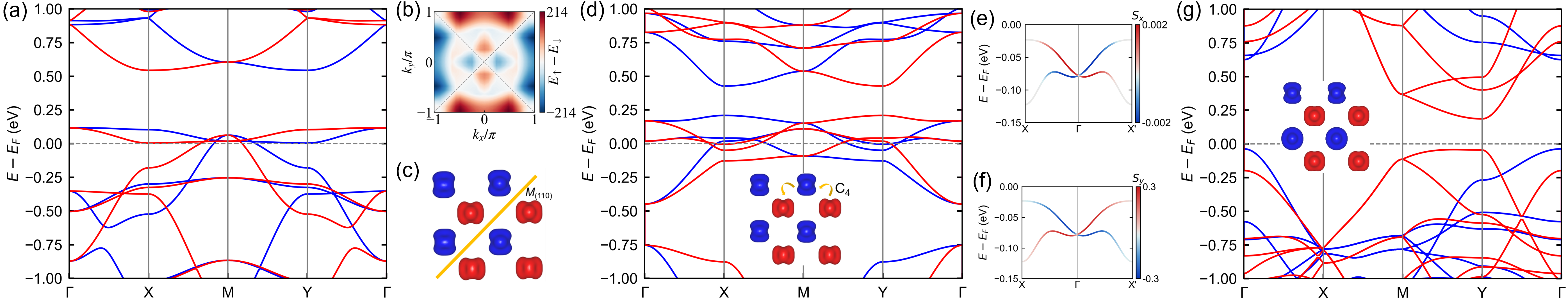}
    \caption{
\textbf{First-principles realization of the three symmetry classes.}
(a--c) Cross-layer Se divacancy, representative of Class I:
spin-resolved bands without SOC, momentum-resolved $d$-wave splitting of the highest occupied band (meV),
and real-space spin density showing the surviving $M_{(110)}$ operation
that relates the opposite-spin V sectors. (d--f) Single Se vacancy,
representative of Class II: compensated bands without SOC and
relativistic bands projected onto $S_x$ and $S_y$. The strongly suppressed $S_x$ and finite $S_y$ components reveal the symmetry-selected p-wave SML.
(g) Non-diagonal O divacancy, representative of Class III: spin-resolved bands
and spin density showing loss of the protected $d$-wave alternation and
the emergence of a ferrimagnetic phase.
}
\label{Fig3}
\end{figure*}

\textit{First-principles verification and emergent responses.—}
The DFT results reproduce the SML for each case. A cross-layer Se
divacancy, representative of Class I, retains compensated spin splitting
despite substantial local reconstruction. Its momentum-space splitting
displays the four-lobed $B_{1g}$ pattern and diagonal nodes predicted by
Eq.~\eqref{eq:main_dwave_splitting}, while the real-space spin density
shows that the opposite-spin V environments remain related by the
$M_{(110)}$ mirror [Figs.~\ref{Fig2}(a,b) and
\ref{Fig3}(a--c)]. Class I defects therefore alter band positions and
splitting amplitudes without changing the underlying magnetic phase. Their
momentum-dependent spin splitting is directly accessible to spin-resolved
photoemission, following its observation in three-dimensional altermagnets
\cite{krempasky2024altermagnetic,reimers2024direct}, while dichroic
imaging can test whether defect-rich regions preserve the parent domain
structure \cite{amin2024nanoscale,hariki2024x}.

\begin{figure*}
    \centering
    \includegraphics[width=0.99\linewidth]{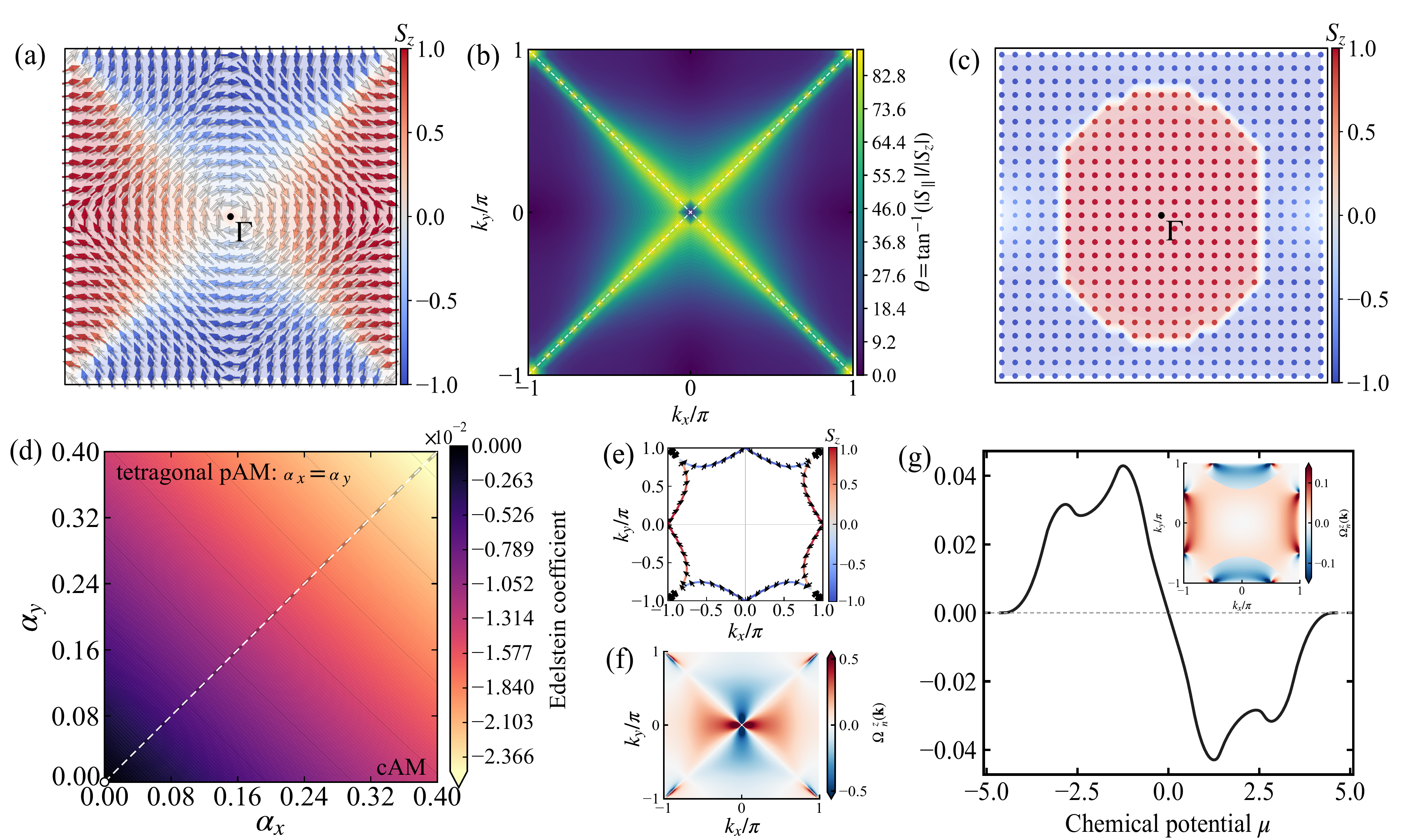}
    \caption{
\textbf{Correspondence between first-principles SML and
model responses.}
(a) SML within DFT of the highest occupied band in the polar altermagnetic phase; arrows denote the in-plane spin and color denotes $S_z$. (b) Corresponding canting angle, enhanced
near the nodes of the $d$-wave exchange. (c) SML within DFT of the highest occupied band of a
Class III defect, showing the loss of the four-lobed sign reversal.
(d) Reduced transverse Edelstein susceptibility as a function of
$\alpha_x$ and $\alpha_y$; the diagonal is the tetragonal polar limit.
(e) Fermi contours with in-plane spin orientation and $S_z$ color,
illustrating the spin--velocity locking responsible for the Edelstein
response. (f) Local Berry curvature of a representative polar band;
its sign-changing distribution integrates to zero. (g) Class III
intrinsic AHC versus Fermi energy, demonstrating the
filling-dependent anomalous Hall response enabled by magnetic-sector inequivalence and SOC.
}
\label{Fig4}
\end{figure*}

The single Se vacancy realizes Class II. Without SOC, its bands remain
compensated and retain the nonrelativistic $d$-wave splitting, confirming
that inversion breaking alone does not destroy altermagnetism. SOC then
produces a strongly component-selective in-plane polarization along
$X$--$\Gamma$--$X'$: the longitudinal component is suppressed, whereas
the transverse component is finite, in agreement with
Eq.~\eqref{eq:main_effective_field}
[Figs.~\ref{Fig2}(c,d) and \ref{Fig3}(d--f)]. Across the Brillouin zone,
the sign-changing out-of-plane component coexists with a circulating
in-plane spin texture, and the spin canting is largest near the nodes of the
$d$-wave exchange [Figs.~\ref{Fig2}(e,f) and \ref{Fig4}(a,b)]. The spin texture in the polar altermagnet follows the trend reported for noncollinear altermagnets \cite{ hu2025spin}: the in-plane component satisfies $S_{i}(\mathbf{k})=-S_{i}(-\mathbf{k})$ with $i=x,y$, giving rise to a p-wave SML which is odd under time-reversal operator\cite{Fukaya_2025}, whereas the out-of-plane component obeys $S_z(\mathbf{k})=S_z(-\mathbf{k})$ that is even under time-reversal operator, corresponding to a quadrupolar SML. The agreement among magnetic symmetry, the projected model, and DFT identifies a relativistic SML composed of p$_y$-, p$_x$- and $d_{x2-y2}$-waves for S$_x$, S$_y$ and S$_z$ spin components, respectively. We define this in short as  
the mixed $d+p$ SML, and this is a signature of a polar altermagnet\cite{liu2025realizing}.

The same $p$-wave spin--momentum locking generates a transverse Edelstein response,
$\delta S_i=\chi_{ij}^{S}E_j$
\cite{edelstein1990spin,gonzalez2024non}. It vanishes at the
centrosymmetric point and becomes finite when the polar SOC is introduced.
The tetragonal limit satisfies
$\chi_{xy}^{S}=-\chi_{yx}^{S}$, while the orthorhombic phase permits two
independent transverse coefficients. The calculated Fermi contours expose
the microscopic origin of the response by correlating carrier velocity
with the in-plane spin orientation [Fig.~\ref{Fig4}(d,e)]. Reversing the
polar distortion reverses the induced spin, providing a symmetry-controlled
experimental discriminator against polarity-even thermal backgrounds\cite{gu2025ferroelectric}.
A corresponding current-induced exchange torque is allowed, although a
quantitative switching efficiency requires a defect-resolved torque
calculation. The polar bands also carry finite local Berry curvature, but the residual
magnetic operations enforce cancellation of its Brillouin-zone integral
[Fig.~\ref{Fig4}(f)]. Together with the $C_{2z}$ prohibition of purely
in-plane second-order charge transport, this identifies the Edelstein
response---rather than a generic nonlinear Hall signal---as the principal
bulk electrical fingerprint of Class II.

Class III is represented by the O divacancy not along a diagonal. Its
momentum-space spin texture no longer exhibits the symmetry-enforced
four-lobed sign reversal and instead develops a ferrimagnetic phase [Figs.~\ref{Fig2}(g,h), \ref{Fig3}(g), and
\ref{Fig4}(c)]. This reconstruction differs fundamentally from the
relativistic canting of Class II: inversion breaking rotates the spin while preserving the underlying $d$-wave exchange, whereas magnetic-sector
inequivalence removes the symmetry that protects the SML.
Residual momentum dependence can remain through hybridization and
lower-symmetry crystal fields.

Loss of magnetic-sector equivalence does not uniquely determine the total
magnetization. Depending on defect-induced occupations, localized states,
and ligand polarization, the relaxed structure may be ferrimagnetic or
globally compensated while its magnetic sectors remain inequivalent. The same symmetry reduction removes the cancellation constraint on the Berry curvature\cite{sorn2025activation}. In the presence of SOC-induced interband mixing, the
representative model develops a finite AHC [Fig.~\ref{Fig4}(g)]. Its pronounced filling dependence, including
sign changes, suggests that electrostatic gating or controlled doping can
tune both the magnitude and direction of the anomalous Hall response\cite{liu2023gate}. Classes I and
II remain intrinsically anomalous Hall silent under their magnetic
symmetries.

The combined symmetry analysis, minimal model, and DFT calculations thus
establish point defects as symmetry-selective control parameters rather
than passive sources of disorder. Preserving inversion and magnetic-sector
equivalence yields a robust $d$-wave altermagnet; removing inversion alone
activates a compensated mixed-spin phase with electrical spin conversion;
and removing the operation relating the magnetic sectors can unlock
anomalous Hall-active magnetism. Defect geometry can therefore select spin-momentum locking, spin--charge-converting, and anomalous  Hall-active states within a single
host.

\textit{Conclusions and outlook.—}
The energetic results show that the symmetry hierarchy is not a consequence
of a fragile magnetic energy balance. Across all defective V$_2$Se$_2$O supercells, the antiparallel magnetic
solution remains lower than the collinear ferromagnetic
reference, while the local V moments remain close to
$2\,\mu_{\rm B}$. The calculated defect formation energies indicate that their formation is feasible compared with other 2D materials\cite{ali2023high}; see Supplemental Material. Defects sharing the same stoichiometry can nevertheless stabilize distinct SML purely through their geometric disposition: configurations such as diagonal versus non‑diagonal O or V divacancies, or intra‑ versus interlayer Se divacancies, relax into inequivalent magnetic symmetries. In contrast, defects of different chemical character collapse onto the same phase whenever they conserve the same sector‑preserving symmetry operations. Thus, thermodynamic cost and electronic functionality separate into independent design axes: defect chemistry governs energetic feasibility, whereas defect geometry, through the global symmetry it maintains, determines the selected magnetic phase.

The same principle extends to other altermagnets beyond V$_2$Se$_2$O. 
Applying the classification
to pentagonal Mn$_4$N$_2$\cite{Zhang2026Mn4N2} and hexagonal 2H-FeBr$_3$\cite{Sodequist2024}  yields the same
symmetry-controlled organization despite their different lattices and
parent altermagnetic harmonics (see Supplemental Material). In
Mn$_4$N$_2$, vacancy patterns that remove the operation relating the
opposite-spin Mn sectors convert the parent $d$-wave SML into a
predominantly ferrimagnetic state. In noncentrosymmetric
2H-FeBr$_3$, the residual symmetry selects a richer sequence: the parent
$i+f$ SML can be retained, reconstructed into mixed $d+s$ SML,
converted into polar $d+p$ or $g+p$ states, or replaced by a predominantly $s$-wave SML. Thus, the universal quantity is not a
particular $d$-, $g$-, or $i$-wave harmonic, but the survival of the
operation relating the magnetic sectors together with the polar or nonpolar character of the residual point group. The host symmetry determines the allowed momentum harmonic, while the geometry of the defect pattern selects the magnetic symmetry that is realized\cite{21z4-c9p2}.
Although the results are presented in terms of impurity lattices with impurities repeated periodically, they remain valid in the limit of large supercells, corresponding to low impurity concentrations. In this limit, the effects on the detailed shape of the band structure are reduced, while the underlying symmetry considerations remain unchanged. For Class III, at low impurity concentrations, each impurity can be regarded as a bound magnetic polaron~\cite{fp4d-grwc}.
Our results establish a materials-independent strategy for using atomically controlled defects to access exchange-split, spin--charge-converting, and anomalous Hall phases within two-dimensional altermagnets.
It has been shown that, under applied strain in pentagonal altermagnets, the symmetry of the SML can remain unchanged, be modified, or be completely broken, resulting in a compensated ferrimagnet\cite{wang2026symmetryselectivestraincontrolspinmomentum}. This behavior is analogous to the changes in relativistic SML induced by impurities in altermagnets. Thus, this classification of modifications to the SML also applies to other perturbations of the altermagnetic phase and provides a general framework for classifying the effects of different external stimuli.

\begin{acknowledgments}
M.B.T. acknowledges the funding support by  Narodowa Agencja Wymiany Akademickiej (NAWA) under the ULAM program with project number BPN/ULM/2025/1/00156/U/00001.
M.B.T. acknowledges the funding support by Iran National Science Foundation (INSF) under project No.4043973. This research was supported by the Foundation for Polish Science project “MagTop” no. FENG.02.01-IP.05-0028/23 co-financed by the European Union from the funds of Priority 2 of the European Funds for a Smart Economy Program 2021–2027 (FENG). We further acknowledge access to the computing facilities of the Interdisciplinary Center of Modeling at the University of Warsaw, Grant g91-1418, g91-1419, g96-1808, g96-1809 and g103-2540 for the availability of high-performance computing resources and support. We acknowledge access to the computing facilities of the Poznan Supercomputing and Networking Center, Grants No. pl0267-01, pl0365-01, pl0807, and pl0471-01.
\end{acknowledgments}

\medskip

\newpage

\appendix

\bibliography{references}
\end{document}